\documentclass[12pt]{article}
\usepackage{euscript,amsmath, amssymb, amsfonts, color}

\newcommand{\supp}{{\rm supp\,}}
\newcommand{\R}{{\mathbb R}}
\newcommand{\G}{{\mathbb G}}
\newcommand{\ad}{d_2^{\mathrm {ad}}}
\newcommand{\K}{{\mathbb K}}
\newcommand{\oC}{{\mathbb C}}
\newcommand{\ooC}{{\mathcal C}}
\newcommand{\di}{{\rm d}}
\newcommand{\p}{{\hbar^2}}
\newcommand{\pp}[2]{\hbar^{#1}}
\newcommand{\oP}{{\cal P}}

\newcommand{\oD}{{\mathbf D}}
\newcommand{\eE}{\EuScript E}
\newcommand{\be}{\begin{equation}}
\newcommand{\ee}{\end{equation}}
\newcommand{\ben}{\begin{equation*}}
\newcommand{\een}{\end{equation*}}

\newcounter{theorem}
\newcommand{\theorem}{\par\refstepcounter{theorem}
           {\bf Theorem \arabic{section}.\arabic{theorem}. }}

\makeatletter \@addtoreset{theorem}{section}

\newcounter{lemma}
\newcommand{\lemma}{\par\refstepcounter{lemma}
           {\bf Lemma \arabic{section}.\arabic{lemma}. }}

\makeatletter \@addtoreset{lemma}{section}

\newcounter{proposition}
\newcommand{\proposition}{\par\refstepcounter{proposition}
           {\bf Proposition \arabic{section}.\arabic{proposition}. }}
\renewcommand\theproposition{\thesection.\arabic{proposition}}
\makeatletter \@addtoreset{proposition}{section}

\newcounter{definition}
\newcommand{\definition}{\par\refstepcounter{definition}
           {\bf Definition \arabic{section}.\arabic{definition}. }}

\makeatletter \@addtoreset{definition}{section}

\begin{document}
\sloppy
\title
 {
            \vspace{1cm}
       \textbf{General form of the deformation of the Poisson superbracket
       on (2,2)-dimensional superspace}
 }
\author
 {
 S.E.~Konstein\thanks{E-mail: konstein@lpi.ru}
~~and~I.V.~Tyutin\thanks{E-mail: tyutin@lpi.ru}
 \thanks{
               This work was supported
               by the RFBR (grants No.~05-01-00996
               (I.T.) and No.~05-02-17217 (S.K.)),
               and by the grant LSS-1578.2003.2.
 }
\\ {\small
               \phantom{uuu}} \\ {\small I.E.Tamm Department of
               Theoretical Physics,} \\
               {\small P. N. Lebedev Physical
               Institute,} \\ {\small 119991, Leninsky Prospect 53,
               Moscow, Russia.} }

\date{}
 \maketitle

\begin{abstract}
{ \footnotesize
Continuous formal deformations of the Poisson superbracket defined
on compactly supported smooth functions on $\R^2$
taking values in a Grassmann algebra $\G^{2}$ are described up to an
 equivalence
transformation.
}
\end{abstract}

\section{Introduction}

In the present paper, we finalize finding
the general form of the $*$-commutator in the
case of a Poisson superalgebra of smooth compactly supported functions on
$\R^{n_+}$ taking
values in a Grassmann algebra $\G^{n_-}$.
We solve this problem for the case $n_+=n_-=2$ and show that in this case
the Poisson superalgebra has an additional deformation comparing with other cases
considered in \cite{deform4} and in
\cite{deform2}.
The proposed analysis is essentially based also on
the results of the papers \cite{n=2} by the authors, where the first and the second
cohomology spaces with coefficients in the adjoint representation of the
Poisson superalgebra are found.

It is interesting to note that anti-Poisson superalgebra based on vector fields
with polynomial coefficients also has an additional deformation
at superdimension $(2,2)$ \cite{LS}.

\subsection{Main definitions}
Here we recall main definitions.

\subsubsection{Poisson superalgebra $\oP$}
Let the field $\K$ be either $\R$ or $\oC$. Let $\EuScript D$ denote the space of
smooth $\K$-valued functions with compact support on $\R^2$.
This space is
endowed by its standard topology.
We set
$$ \mathbf D= \EuScript
D\otimes \G^{2},\quad \mathbf E=
C^\infty(\R^{2})\otimes \G^{2},
$$
where $\G^{2}$ is the Grassmann
algebra with $2$ generators.
The generators of
the Grassmann algebra (resp., the coordinates of the space $\R^{2}$) are
denoted by $\xi^\alpha$, $\alpha=1,2$ (resp., $x^i$, $i=1,2$).
We shall also use collective variables $z^A$ which are equal to $x^A$
for $A=1,2$ and are equal to $\xi^{A-2}$ for
$A=3,4$.
  The spaces $\mathbf D$ and $\mathbf
E$ possess a natural grading
which is determined by that of the Grassmann algebra. The parity of an
element $f$ of these spaces is denoted by $\varepsilon(f)$.

The Poisson bracket is
defined by the relation
\begin{equation}
\{f,g\}(z)=f(z)\frac{\overleftarrow{\partial}}{\partial z^A}\omega^{AB}
\frac{\partial}{\partial z^B}g(z)=
- (-1)^{\varepsilon(f)\varepsilon(g)} \{g,f\}(z),\label{3.0}
\end{equation}
where
the symplectic metric $\omega^{AB}=-(-1)^{\varepsilon_A
\varepsilon_B}\omega^{BA}$ is a constant invertible matrix.  For
 definiteness, we choose it in the form
$$
\omega^{AB}=
\left(\begin{array}{cc}\omega^{ij}&0       \\
0&\lambda_\alpha\delta^{\alpha\beta}\end{array}\right),\quad
\lambda_\alpha=\pm1,\ i,j=1,2,\ \alpha,\beta=1,2\,,
$$
 where
$\omega^{ij}$ is the canonical symplectic form (if $\K=\oC$, then one can
choose $\lambda_\alpha=1$).
The Poisson superbracket
satisfies the Jacobi identity
\begin{equation}
 (-1)^{\varepsilon(f)\varepsilon(h)} \{f,\{g,h\}\}(z)+
 \hbox{cycle}(f,g,h)= 0,\quad f,g,h\in \mathbf E.  \label{3.0a}
\end{equation}
By Poisson superalgebra
$\mathcal P$,
we mean the space $\mathbf D$ with the Poisson
bracket~(\ref{3.0}) on it.  The relations~(\ref{3.0}) and~(\ref{3.0a}) show
that this bracket indeed determines a Lie superalgebra structure on $\mathbf
D$.

The integral on $\mathbf D$ is defined by the relation
$$
\bar
f\stackrel{\mathrm{def}}{=}\int \di z\, f(z)= \int_{\R^2}\di x\int
\di\xi\, f(z),
$$
where the integral on the Grassmann algebra is normed by
the condition $\int \di\xi\, \xi^1\xi^2=1$.

\subsubsection{Cohomologies of $\oP$}
The space $\ooC_p(L,V)$ of
$p$-cochains consists of all multilinear super-antisymmetric mappings
from $L^p$ to $V$.

Nilpotent differentials $d_1^{\mathrm {ad}}$ mapping $\ooC_1(\mathbf E,\mathbf E)$
to $\ooC_2(\mathbf E,\mathbf E)$
and $d_2^{\mathrm {ad}}$ mapping $\ooC_2(\mathbf E,\mathbf E)$
to $\ooC_3(\mathbf E,\mathbf E)$
have the form
\begin{eqnarray*}
&&d_{1}^{\mathrm{ad}}M_{1}(z|f,g)=
\{f(z),M_{1}(z|g)\}-
(-1)^{\varepsilon(f)\varepsilon(g)}\{g(z),M_{1}(z|f)
\}-M_{1}(z|\{f,g\}),\\
&&d_{2}^{\mathrm{ad}}M_{2}(z|f,g,h)=\\
&&\quad\quad
=(-1)^{\varepsilon(f)\varepsilon(h)}\left[(-1)^{\varepsilon(f)\varepsilon(h)}
(\{f(z),M_{2}(z|g,h)\}
+M_{2}(z|f,\{g,h\}))+\mbox{cycle}(f,g,h)\right].
\end{eqnarray*}

Let
$\Lambda\in
C^\infty({\mathbb R})$ be a function such that the function
$\frac d {dx} \Lambda$ has a compact support,
$\Lambda(-\infty) = 0$, and $\Lambda( + \infty) = 1$,
\begin{eqnarray}
\Theta(x|f) & \stackrel {def} = &\int du
\delta(x_1-u_1)\theta(x_2-u_2)f(u),                          \\
\tilde{\Delta}(x^{1}|f)& \stackrel {def} = & \int du\delta (x^{1}-u^{1})f(u).
\end{eqnarray}

It is easily to prove that
linear mapping $\Xi$,
\ben
\Xi (x|f) \stackrel {def} = \Theta (x|f)-\tilde{\Delta}(x^{1}|f)\Lambda (x^{2}),
\een
maps $\mathbf D$ to $\mathbf D$.

Let
$\mathbf{Z}=\mathbf{D} \oplus {\cal C}_{\mathbf{E}}(\mathbf{D})$,
where ${\cal C}_{\mathbf{E}}(\mathbf{D})$ is a centralizer of
$\mathbf{D}$ in $\mathbf{E}$.
Evidently, ${\cal C}_{\mathbf{E}}(\mathbf{D}) \simeq \K$.

The following Theorem is proved in \cite{n=2}:

\theorem\label{th2}
{
\it
\label{p23}

Let the bilinear mappings $m_1$,
$m_3$,
 and $m_5$ from
 $\mathbf D^2$ to $\mathbf D$ be defined by the relations
\begin{eqnarray}
m_1(z|f,g)& =
&f(z)\!\left(\frac{\overleftarrow{\partial}}{\partial z^A} \omega^{AB}
\frac{\partial}{\partial z^B} \right)^3\!g(z),       \label{m1}
\\
m_3(z|f,g)& = & \eE_z f(z)\bar g -
(-1)^{\varepsilon(f)\varepsilon(g)}  \eE_z g(z)\bar f,       \label{m3}
\\
m_{5}(z|f,g)&=&m_{51}(z|f,g)+m_{52}(z|f,g)+m_{53}(z|f,g)+m_{54}(z|f,g),
\label{m5}
\\
m_{51}(z|f,g)&=&f(z)\Delta (x|g)-\Delta (x|f)g(z),       \nonumber
\\
m_{52}(z|f,g)&=&2\Theta (x|f)\partial _{2}g(z)-
2\partial _{2}f(z)\Theta (x|g),                          \nonumber
\\
m_{53}(z|f,g)&=&-2\{f(z),\Psi (x|g)\}+2\{g(z),\Psi (x|f)\},   \nonumber
\\
m_{54}(z|f,g)&=&\Xi (x|\partial _{2}fg-f\partial _{2}g),
                                                      \nonumber
\end{eqnarray}
where
\begin{eqnarray}
\eE_z & \stackrel {def} = & 1-\frac 1 2 z \partial_z,    \label{Ez}
\\
\Delta (x|f) & \stackrel {def} = &\int du \delta(x-y) f(u),        \nonumber
\\
\tilde{\Theta}(x^{1}|f)& \stackrel {def} = &
\int du\theta (x^{1}-y^{1})f(u),                                     \nonumber
\\
\Psi (x|f)& \stackrel {def} = &\tilde{\Theta}(x^{1}|f)\Lambda (x^{2}), \nonumber
\end{eqnarray}
and
$z = (x^1,x^2,\xi^1,\xi^2)$,
$u = (y^1,y^2,\eta^1,\eta^2)$.

Let $V^3$ be the subspace of ${\cal C}_2(\mathbf D,\mathbf
D)$ generated by the cocycles $m_1$, $m_3$ and $m_5$.

Then there is a natural isomorphism
$V^3\oplus ({\raise2pt\hbox{$\mathbf E$}}\big/{\raise-2pt\hbox{$\mathbf Z$}})
\simeq H^2_{\mathrm{ad}}$ taking $(M_2,T)\in
V^3\oplus ({\raise2pt\hbox{$\mathbf E$}}\big/{\raise-2pt\hbox{$\mathbf Z$}})$
to the cohomology
class determined by the cocycle
\ben
 M_2(z|f,g)+m_\zeta(z|f,g),
\een
where
\be\label{m_zeta}
m_\zeta(z|f,g)= \{\zeta(z),f(z)\} \bar{g} -
  \{\zeta(z),g(z)\} \bar{f},
\ee
and even function $\zeta\in \mathbf E$
belongs to the equivalence class $T$.
}


If $\zeta^1\in\mathbf E$ and $\zeta^2\in\mathbf E$
belong to the same equivalence class of
${\raise2pt\hbox{$\mathbf E$}}\big/{\raise-2pt\hbox{$\mathbf Z$}}$ then
${\zeta^1}-{\zeta^2}=(\zeta^1-\zeta^2)^D+c_{{\zeta^1}-{\zeta^2}}$
where $(\zeta^1-\zeta^2)^D \in \oD$ and $c_{{\zeta^1}-{\zeta^2}}\in\K$.
In this case $m_{\zeta^1}-m_{\zeta^2}$ is coboundary because
$m_{\zeta^1}-m_{\zeta^2}=d_1^{\mathrm {ad}}m_1$, where
$m_1(f)=(\zeta^2-\zeta^1)^D\bar f\in {\cal C}_1(\oD,\oD)$.

Thus, the general solution of the equation
$d_{2}^{\mathrm{ad}}M_{2}(z|f,g,h)=0$
has the form
\be\label{cohom}
M_{2}(z|f,g)=c_{1}m_{1}(z|f,g)+c_{3}m_{3}(z|f,g)+c_{5}m_{5}(z|f,g)+m_\zeta(z|f,g)+
d_{1}^{\mathrm{ad}}b^D(z|f,g),
\ee
where
$b^D(f,g) \in \oD$ and
$\zeta\in{\raise2pt\hbox{$\mathbf E$}}\big/{\raise-2pt\hbox{$\mathbf Z$}}$.

\subsubsection{Deformations of topological Lie superalgebras}

In this paper we find
 formal deformations of $\oP$.
Let  $\K[[\p]]$ be the ring of
formal power series in $\p$ over $\K$, and $\oP[[\p]]$ be the $\K[[\p]]$-module
of formal power series in $\p$ with coefficients in $\oP$.

A (continuous) formal deformation of $\oP$ is by definition a
$\K[[\p]]$-bilinear separately continuous Lie superbracket $C(\cdot,\cdot)$
on $\oP[[\p]]$ such that $C(f,g)=\{f,g\} \mod \p$ for any $f,g\in \oP[[\p]]$.
Obviously, every formal deformation $C$ is expressible in the form
\begin{equation}\label{1} C(f,g)=\{f,g\}+\p
C_1(f,g)+\pp{4}{2}C_2(f,g)+\ldots,\quad f,g\in L,
\end{equation}
where $C_j$
are separately continuous skew-symmetric bilinear mappings from $\oD\times \oD$
to $\oD$ (2-cochains with coefficients in the adjoint representation of $\oP$).
We will denote the bilinear form $\{f,g\}$ as $C_0(f,g)$.
Formal deformations $C^1$ and $C^2$ are called equivalent if there is a
continuous $\K[[\p]]$-linear operator $T=\mathrm {id}+\p T_1+\pp{4}{2} T_2+ ...\ :
\oD[[\p]]\to \oD[[\p]]$ such that
$TC^1(f,g)=C^2(T f,Tg)$, $f,g\in \oD[[\p]]$.

\section{Formulation of the results}

For any $\varkappa\in \K[[\hbar]]$, such that $c_1\stackrel {def}=\frac 1 6
\p\varkappa^2\in
\p\K[[\p]]$, the
Moyal-type superbracket
\begin{equation}\label{2} {\cal
M}_\varkappa(z|f,g)=\frac{1}{\hbar\varkappa}f(z)\sinh
\left(\hbar\varkappa\frac{\overleftarrow{\partial}}{\partial z^A}\omega^{AB}
\frac{\partial}{\partial z^B}\right)g(z)
\end{equation}
is skew-symmetric and
satisfies the Jacobi identity
and, therefore, gives a deformation of the
initial Poisson algebra.
For $\zeta\in {\raise2pt\hbox{$\mathbf \p E[[\p]]$}}
\big/{\raise-2pt\hbox{$\mathbf Z[[\p]]$}}$,
$c_1\in \p\K[[\p]]$, we set
\begin{align}
&{\cal N}_{c_1,\zeta}(z|f,g)= {\cal
M}_\varkappa(z|f-\zeta\bar{f},g-\zeta\bar{g}),\nonumber
\end{align}

If $\zeta^1$ and $\zeta^2$
belong to the same equivalence class of
${\raise2pt\hbox{$\mathbf E[[\p]]$}}\big/{\raise-2pt\hbox{$\mathbf Z[[\p]]$}}$
then
the deformations
${\cal N}_{c_1,\zeta^1}$ and ${\cal N}_{c_1,\zeta^2}$ are equivalent \cite{deform4},
\cite{deform2}.

Now we can formulate the main result of the present paper.

\medskip\noindent
\theorem\label{th1}{\it
Let bilinear forms $m_1$, $m_3$, $m_\zeta$ and $m_5$ are defined as in Theorem \ref{th2}.
Let bilinear forms $\sigma_3$, $\sigma_\zeta$ and $B_\zeta$ are defined as follows:

\begin{eqnarray*}
\sigma _{3}(f,g) &=&4\widetilde{\Theta }(g)  \left( \Xi (f)
\partial _{2}\Lambda -\Xi (f  \partial _{2}\Lambda )\right)
-4\widetilde {\Theta }(f)  \left( \Xi (g)  \partial _{2}\Lambda -\Xi (g
\partial _{2}\Lambda )\right),
\\
\sigma _{\zeta }(f,g) &=&A_\zeta(f)  \overline{g}-A_\zeta(g)  \overline{f},
\\
&&A_\zeta(f) =f  \Delta (\zeta )-\Delta (f)  \zeta +\Xi (\partial
_{2}f  \zeta -f  \partial _{2}\zeta )+2\partial _{1}\zeta
\partial _{2}\Lambda   \widetilde{\Theta }(f)+
\\
&&\ \ \ \ \ \ \ \ \ \ \
+2\partial _{2}\zeta
\Lambda   \Xi (f)-2\partial _{2}f
\int \theta(x_2-y_2)\theta(y_2)\delta(x_1-y_1) \zeta(u)){\mathrm d}u,
\\
B_\zeta(f,g) &=&4\left( \widetilde{\Theta }(f)  \overline{g}-
\widetilde{\Theta }(g)
  \overline{f}\right)
  \left[ \Xi
(\zeta   \partial _{2}\Lambda )-\partial _{2}\Lambda
\int \theta(x_2-y_2)\theta(y_2)\delta(x_1-y_1) \zeta(u)){\mathrm d}u \right].
\end{eqnarray*}

Then every
continuous formal deformation of the Poisson superalgebra $\mathcal P$ is
equivalent either to the superbracket
\be
\label{dkappa}
{\cal N}_{c_1,\zeta}(z|f,g),
\ee
or
to the superbracket
\be
\label{dm3}
C^{(3)}_{c_3,\zeta}(z|f,g)=\{f(z),g(z)\}+m_{\zeta}(z|f,g)+ c_3 m_3(z|f,g),
\ee
or
to the superbracket
\be
\label{dm5}
C^{(5)}_{c_5,\zeta}(z|f,g) \! = \!
\{f(z),g(z)\}+m_{\zeta}(z|f,g)+ c_5[m_5(z|f,g)-\sigma_\zeta(z|f,g)]
+c_5^2[B_\zeta(z|f,g)-\sigma_3(z|f,g)],
\ee
where
$\zeta\in {\raise2pt\hbox{$\p\mathbf
E[[\p]]$}}\big/{\raise-2pt\hbox{$\mathbf Z[[\p]]$}}$ is even and
$c_1,\,c_3,\,c_5\in \p\K[[\p]]$.
}

We will use the decomposition
\begin{eqnarray}
c_1=\sum_{s=1}^\infty c_{1s} \hbar^{2s}\,,\quad
c_3=\sum_{s=1}^\infty c_{3s}\hbar^{2s}\,,\quad
c_5=\sum_{s=1}^\infty c_{5s}\hbar^{2s}\,,\quad
\zeta=\sum_{s=1}^\infty \zeta_{s}\hbar^{2s}
\end{eqnarray}
and partial sums
\begin{eqnarray}
c_{1[k]}=\sum_{s=1}^k c_{1s} \hbar^{2s}\,,\quad
c_{3[k]}=\sum_{s=1}^k c_{3s}\hbar^{2s}\,,\quad
c_{5[k]}=\sum_{s=1}^k c_{5s}\hbar^{2s}\,,\quad
\zeta_{[k]}=\sum_{s=1}^k \zeta_{s}\hbar^{2s}
\end{eqnarray}

\section{Jacobiators}
Let $p$, $q$ be bilinear forms.
Jacobiators are defined as follows:
\begin{eqnarray*}
J(p,q)&\stackrel {def} = &
(-1)^{\varepsilon(f)\varepsilon(h)}
\left((-1)^{\varepsilon(f)\varepsilon(h)}
p(f,q(g,h))+q(f,p(g,h))+\mbox{cycle}(f,g,h)\right),\\
J(p,p)&\stackrel {def} = & (-1)^{\varepsilon(f)\varepsilon(h)}\left(
(-1)^{\varepsilon(f)\varepsilon(h)}p(f,p(g,h))+\mbox{cycle}(f,g,h)\right).
\end{eqnarray*}
Evidently, $J(p,q)\in {\cal C}_3(\oD , \oD)$,
if $p,q\in {\cal C}_2(\oD , \oD)$.

If $m_0(f,g)=\{f,g\}$ then $J(p,m_0)=d_2^{\mathrm {ad}}p$.

Sometimes we will use notations
$J_{ab}\stackrel {def}= J(m_a, m_b)$ for Jacobiators of coboundaries.

One can derive the following relations by direct calculations:

\begin{eqnarray}
J(m_3,m_\zeta)=J(m_3,m_3)=J(m_{\zeta^1 },m_{\zeta^2 })&=&0,
\nonumber\\
J(m_{\zeta },m_{5}) &=&\ad\sigma _{\zeta },
\nonumber  \\
J(m_{5},m_{5}) &=&\ad\sigma _{3},
\nonumber  \\
J(m_{5},\sigma _{3})=
J(m_{\zeta^1 },\sigma _{\zeta^2 })=
J(\sigma _{3},\sigma _{3})=
J(\sigma _{\zeta^1 },\sigma _{\zeta^2 })&=&0,
\nonumber  \\
J(m_{\zeta },\sigma _{3})+J(m_{5},\sigma _{\zeta }) &=&\ad B_\zeta,
\nonumber  \\
J(\sigma _{3},\sigma _{\zeta })+J(m_{5},B_\zeta) &=&0,
\nonumber  \\
J(m_{\zeta^1 },B_{\zeta^2}) =
J(\sigma _{\zeta^1 },B_{\zeta^2}) =
J(\sigma _{3},B_\zeta)=
J(B_{\zeta^1},B_{\zeta^2}) &=&0.             \label{j5}
\end{eqnarray}

Besides,
\begin{eqnarray*}
J(m_1,m_\zeta) & = & d_2^{\mathrm {ad}} \mu_\zeta,\\
J(m_1,m_5) & = & d_2^{\mathrm {ad}} \nu + I_{1,5},
\end{eqnarray*}
\be\label{zeta}
J(\mu_{\zeta^1},\mu_{\zeta^2}) =
J(m_{\zeta^1},\mu_{\zeta^2})=0,
\ee
where
\begin{eqnarray*}
\mu_{\zeta }(z|f,g) & = & m_1(f,\zeta)\bar g - m_1(g,\zeta)\bar f,\\
\nu(z|f,g)& = &
\partial _{1}^{2}\Xi (x|f\partial _{2}^{3}g)
 +2\partial _{2}^{3}f(z)\partial _{1}^{2}\Theta(x|g)
 +2m_{1}(z|f,\Psi (g))-
 (-1)^{\varepsilon(f)\varepsilon(g)}(f\leftrightarrow g),
\end{eqnarray*}
and $I_{1,5}$ is local form, which is described in Appendix \ref{AppI}.

\proposition \label{prop2}
{\it Local trilinear form $I_{1,5}$ belongs to nontrivial cohomology class, i.e.
it can not be represented in the form
$I_{1,5} = d_2^{\mathrm {ad}} S$ with some bilinear form $S$.
Equivalently, the equation
$
d_2^{\mathrm {ad}} S =\alpha I_{1,5}
$
has a solution if and only if $\alpha=0$.
}

The proof presented in Appendix \ref{App2}.


\section{The proof of Theorem \ref{th1}}

First of all let us note that the bilinear forms $C^{(3)}_{c_3,\zeta}$ and
$C^{(5)}_{c_5,\,\zeta}$
defined in Theorem \ref{th1} are  indeed
the deformations.

This statement follows from the relations (\ref{j5}).

Further we solve the equation
\be\label{main}
J(C,C)=0
\ee
 by its decomposition in orders of $\p$.

As $C=\sum_0^\infty \hbar^{2i} C_i$, $C_0(f,g)=\{f,g\}$, this decomposition gives
\be\label{Amain}
\ad C_s + \sum _{1\leq p\leq s/2} J(C_p \,,C_{s-p})=0.
\ee
For $s=1$, Eq. (\ref{Amain}) reduces to
\begin{equation*}
d_{2}^{\mathrm{ad}}C_{1}(z|f,g,h)=0,
\end{equation*}
and has the solution described by
Exp. (\ref{cohom}),
\begin{eqnarray*}
&&C_{1}(z|f,g)=
c_{11}m_{1}(z|f,g)+c_{31}m_{3}(z|f,g)+c_{51}m_{5}(z|f,g)+ \\
&&\,+m_{\zeta _{1}}(z|f,g)+d_{1}^{\mathrm{ad}}b_{1}^{D}(z|f,g),
\end{eqnarray*}
and after similarity transformation eliminating $b_1^D$,%
\footnote
{
The similarity transformation with $T=\mathrm {id}+\p T_1 +\,...$,
where $T_1(f)=- b_1^D(f)$, eliminates $b_1^D$. Below we will
omit all $b_i^D$, which can be eliminated by similarity transformations.
}
$C_1$ acquires the form
\begin{equation*}
C_{1}(z|f,g)=
c_{11}m_{1}(z|f,g)+c_{31}m_{3}(z|f,g)+c_{51}m_{5}(z|f,g)+m_{\zeta
_{1}}(z|f,g).
\end{equation*}

If $c_{11}=c_{31}=c_{51}=0$, then the next order of Eq. (\ref{main}) gives
$\ad C_2=0$ due to relations (\ref{j5}), and
\begin{equation*}
C_{2}(z|f,g)=
c_{12}m_{1}(z|f,g)+c_{32}m_{3}(z|f,g)+c_{52}m_{5}(z|f,g)+m_{\zeta
_{2}}(z|f,g).
\end{equation*}

Evidently, if
$C_i=m_{\zeta_i}$
for $1\le i \le k-1$
then the next order of Eq. (\ref{main}) gives
$\ad C_k=0$.

Let $k\geq 1$ be such, that
\be
C_i=m_{\zeta_i}
\ee
for $1\le i \le k-1$
and
\be
C_k=c_{1k} m_1 + c_{5k} m_5 + c_{3k} m_3 + m_{\zeta_k},
\ee
where at least one of the constants $c_{1k}$, $c_{3k}$ and $c_{5k}$
is not zero.

If such $k$ does not exist, then the deformation has the form
$C(f,g)=\{f,g\}+ m_{\zeta(\p)}(z|f,g)$, which is the partial form of
any of the deformations (\ref{dkappa}), (\ref{dm3}), (\ref{dm5}).

Let such $k$ does exist.

\lemma\label{lem}
{\it
Only one of the constants $c_{1k}$, $c_{3k}$ and $c_{5k}$
is nonzero.
}

The statement of this Lemma follows from Propositions \ref{3k5k}, \ref{1k5k}
and \ref{3k}, which are formulated and proved below.

Consider the equation (\ref{main})
in the degrees $k+1$, $k+2$, ... , $2k-1$ of decomposition
in $\p$ taking in account (\ref{j5}) and (\ref{zeta}):
\ben
\ad (C_{k+1}+c_{5k}\sigma_{\zeta_1,\, 5}+c_{1k}\mu_{\zeta_1})=0,
\een
which implies
\ben
C_{k+1}=-c_{5k}\sigma_{\zeta_1}-c_{1k}\mu_{\zeta_1}+
c_{1,k+1} m_1 + c_{5,k+1} m_5 + c_{3,k+1} m_3 + m_{\zeta_{k+1}}.
\een

In the same way, for $k < s < 2k$:
\ben
 C_{s} = c_{1s}m_1 +c_{3s}m_3+c_{5s}m_5
-\sum_{u+v=s}\hbar^{2u+2v} c_{5u}\sigma_{\zeta_v}
-\sum_{u+v=s}\hbar^{2u+2v} c_{1u}\mu_{\zeta_v}.
\een

Represent $C$ in the form
\ben
C={\cal N}_{c_{1[2k-1]},\zeta_{[2k-1]}}+c_{5[2k-1]} m_5 + c_{3[2k-1]} m_3-
 c_{5[2k-1]}\sigma_{\zeta_{[2k-1]}}+\hbar^{2k}C_{2k}^\prime+O(\hbar^{2k+2}),
\een
where $C_{2k}^\prime=C_{2k}-
({\cal N}_{c_{1[2k-1]},\zeta_{[2k-1]}}-c_{5[2k-1]}\sigma_{\zeta_{[2k-1]}})|_{2k}$,
and the notation $F(\p)|_s$ extracts the coefficient at $\hbar^{2s}$:
$F(\p)=\sum\hbar^{2s} F(\p)|_s$.

Consider the terms of $2k$-th degree in (\ref{main}):
\be\label{25}
\ad D_{2k}
+c_{1k}c_{3k}J_{1,3}+c_{1k}c_{5k}I_{1,5}
+c_{3k}c_{5k}J_{3,5}
=0
\ee
where
\ben
D_{2k}=
C_{2k}^\prime
+c_{5k}^2 \sigma_3
+\sum_{p+q=2k}c_{5p}\sigma_{\zeta_q}
+c_{1k}c_{5k}\nu \in {\cal C}_2(\oD,\oD).
\een

\proposition\label{3k5k}
$c_{3k}c_{5k}=0$.

{\bf Proof.}
Consider Eq. (\ref{25}) in the following domain
\begin{eqnarray*}
&&\,[z\cup \mathrm{supp}(h)]\cap \lbrack \mathrm{supp}(f)\cup \mathrm{supp}
(g)]=\mathrm{supp}(f)\cap \mathrm{supp}(g)=\varnothing , \\
&&f(z)=f_{1}(x^{1})f^{\prime }(x^2,\xi^1,\xi^2),\;g(z)=g_{1}(x^{1})g^{\prime
}(x^2,\xi^1,\xi^2).
\end{eqnarray*}
Here the short notation $z\cap U$ or
$z\cup U$ ($z=(x,\xi)$) means
$V_x\cap U$ or
$V_x\cup U$ correspondingly where $V_x$ is some vicinity  of $x$.
Define $\overline {f^{\prime }}$ as
$\overline {f^{\prime }}=\int f^{\prime }(x^2,\xi^1,\xi^2)\mathrm dx^2 \mathrm d\xi^2
\mathrm d\xi^1$.

In this domain, we have%
\footnote
{
Here and below the sign $\hat {\  }$ over form means that we consider the
restriction of the form on the domain under consideration.
}
\begin{eqnarray*}
&&\hat{J}_{1,3}=\hat{I}_{1,5}=0,\\
&&\hat{J}_{3,5}(z|f,g,h)=\{h(z),\hat{\sigma}_{2}(x|f,g)\},
\end{eqnarray*}
where
\begin{eqnarray*}
&&\sigma _{2}(x|f,g)=\sigma
_{2|1}(x|f,g)+\sigma _{2|2}(x|f,g), \\
&&\sigma _{2|1}(x|f,g)=[\tilde{\Theta}(x^{1}|g)\bar{f}-\tilde{\Theta}
(x^{1}|f)\bar{g}]x^{2}\partial _{2}\Lambda (x^{2})\in {\cal C}_2(\oD,\oD), \\
&&\sigma _{2|2}(x|f,g)=[\tilde{\Delta}(x^{1}|g)\bar{f}-\tilde{\Delta}
(x^{1}||f)\bar{g}]x^{1}\Lambda (x^{2}) \notin {\cal C}_2(\oD,\oD),
\end{eqnarray*}
and Eq. (\ref{25}) takes the form
\begin{eqnarray}
&&\{h(z),(\hat{D}_{2k}^\prime+c_{3k}c_{5k}\hat{\sigma}_{2|2})(x|f,g)\}=0,
\label{26}
\\
&&
D_{2k}^\prime=
D_{2k}+ c_{3k}c_{5k}{\sigma}_{2|1}
\in {\cal C}_2(\oD,\oD).\notag
\end{eqnarray}
It follows from Eq. (\ref{26})
\begin{equation*}
\partial _{2}(\hat{D}_{2k}^\prime+c_{3k}c_{5k}\hat{\sigma}_{2|2})(x|f,g)=0,
\end{equation*}
which gives after integrating over $x^{2}$
\begin{equation*}
c_{3k}c_{5k}[g_{1}(x^{1})\overline{g^{\prime }}\bar{f}-f_{1}(x^{1})\overline{
f^{\prime }}\bar{g}]=0
\end{equation*}
and so
$c_{5k}c_{3k}=0$.
\hskip 5mm
$\blacksquare$

\proposition\label{1k5k}
{\it
If $c_{3k}=0$ then $c_{1k}c_{5k}=0$.
}

{\bf Proof.}
If $c_{3k}=0$, then Eq. (\ref{25}) acquires the form
\ben
\ad D_{2k}
+c_{1k}c_{5k}I_{1,5}=0
\een
and $c_{1k}c_{5k}=0$
according to Proposition \ref{prop2}.
\hskip 5mm
$\blacksquare$

\proposition\label{3k}
{\it
If $c_{3k}\ne 0$ then $c_{1k}=c_{5k}=0$.
}

{\bf Proof.}
If $c_{3k}\ne 0$, then according to Proposition \ref{3k5k}
$c_{5k}=0$ and Eq. (\ref{25}) acquires the form
\be\label{???}
\ad C_{2k}
+c_{1k}c_{3k}J_{1,3}=0
\ee
It is shown in \cite{Cohom}, \cite{deform4}
that Eq. (\ref{???}) has a solution if and only if
$c_{1k}c_{3k}=0$.
So $c_{1k}=0$.
\hskip 5mm
$\blacksquare$

Thus, we have proved Lemma \ref{lem}.

Now, it remains to consider 3 cases separately:
\begin{enumerate}
\item $c_{1k}\ne 0$, $c_{3k}=c_{5k}=0$,
\item $c_{3k}\ne 0$, $c_{1k}=c_{5k}=0$,
\item $c_{5k}\ne 0$, $c_{1k}=c_{3k}=0$.
\end{enumerate}
and to show that these cases lead to deformations
${\cal N}_{c_{1},\zeta}$,
$C^{(3)}_{c_3,\zeta}$ and $C^{(5)}_{c_{5},\zeta}$ correspondingly.

\proposition\label{def1}
{\it
Let $ c_{1k}\ne 0$, $ c_{3k}=c_{5k}=0$ and let
$l>k$ be such that
\ben
C_s=m_{\zeta_s}
\een
for $ s < k$,
\ben
C={\cal N}_{c_{1[s]},\zeta_{[s]}}+O(\hbar^{2s+2})
\een
for $k \leq s < l$,
\ben
C={\cal N}_{c_{1[l]},\zeta_{[l]}}+O(\hbar^{2l+2})+\hbar^{2l}(c_{3l}m_3+c_{5l}m_5).
\een
Then $c_{3l}=c_{5l}=0$.
}

{\bf Proof.}
Acting in the same manner as for deriving the Eq. (\ref{25}), we obtain that
the terms of degree $l+k$ in the decomposition of Eq. (\ref{main})
give the equation:
\be\label{c1}
\ad (C_{k+l}+C_{k+l}^\prime)
+ c_{1k}c_{3l}J_{1,3}+c_{1k}c_{5,l}J_{1,5}+c_{5,l}J_{5,\zeta_k}=0,
\ee
where $C_{k+l}^\prime \in {\cal C}_2(\oD,\oD)$ is some bilinear form
built from known lower order terms.
Eq. (\ref{c1}) is the partial case of Eq. (\ref{25}) and gives
$c_{1k}c_{3l}=c_{1k}c_{5l}=0$, which implies $c_{3l}=c_{5l}=0$.
\hskip 5mm
$\blacksquare$

In such a way this case leads to the deformation ${\cal N}_{c_{1},\zeta}$.

\proposition\label{def3}
{\it
Let $ c_{3k}\ne 0$, $ c_{1k}=c_{5k}=0$ and let
$l>k$ be such that
\ben
C_s=m_{\zeta_s}
\een
for $ s < k$,
\ben
C_s=c_{3s}m_3+m_{\zeta_s}
\een
for $k \leq s < l$,
\ben
C_l=c_{3l}m_3+m_{\zeta_l}+c_{1l}m_1+c_{5l}m_5.
\een
Then $c_{1l}=c_{5l}=0$.
}

{\bf Proof.}

Consider the terms of degree $l+k$ in the decomposition of Eq. (\ref{main}).
\ben
\ad C_{k+l} + c_{3k}c_{1l}J_{1,3}+c_{3k}c_{5,l}J_{3,5}+
c_{5,l}J_{5,\zeta_k}+c_{1l}J_{1,\zeta_k}=0.
\een
or, equivalently,
\be\label{c3}
\ad (C_{k+l}+
c_{5,l}\sigma_{\zeta_k}+c_{1l}\mu_{\zeta_k} )
 + c_{3k}c_{1l}J_{1,3}+c_{3k}c_{5,l}J_{3,5}=0.
\ee
Eq. (\ref{c3}) is the partial case of Eq. (\ref{25}) and gives
$c_{3k}c_{1l}=c_{3k}c_{5l}=0$, which implies $c_{1l}=c_{5l}=0$.
\hskip 5mm
$\blacksquare$

In such a way this case leads to the deformation $C^{(3)}_{c_3,\zeta}$.

\proposition\label{def5}
{\it
Let $ c_{5k}\ne 0$, $ c_{1k}=c_{3k}=0$ and let
$l>k$ be such that
\ben
C_s=m_{\zeta_s}
\een
for $ s < k$,
\ben
C=C^{(5)}_{c_{5[s]},\zeta_{[s]}}+O(\hbar^{2s+2})
\een
for $k \leq s < l$,
\ben
C=C^{(5)}_{c_{5[l]},\zeta_{[l]}}
+ \hbar^{2l}(c_{1l} m_1 + c_{3l}m_3)
+ O(\hbar^{2l+2}).
\een
Then $c_{1l}=c_{3l}=0$.
}

{\bf Proof.}

Consider the terms of degree $l+k$ in the decomposition of Eq. (\ref{main}):
\be\label{c5}
\ad (C_{k+l}+C_{k+l}^\prime )
 + c_{5k}c_{1l}J_{1,5}+c_{5k}c_{3,l}J_{3,5}=0,
\ee
where $C_{k+l}^\prime \in {\cal C}_2(\oD,\oD)$ is some bilinear form
built from known lower order terms.

 Eq. (\ref{c1}) is the partial case of Eq. (\ref{25}) and gives
$c_{5k}c_{1l}=c_{5k}c_{3l}=0$, which implies $c_{1l}=c_{3l}=0$.
\hskip 5mm
$\blacksquare$

In such a way this case leads to the deformation $C^{(5)}_{c_{5},\zeta}$.

Thus, the Theorem \ref {th1} is proved.

\bigskip


\setcounter{equation}{0} \def\theequation{A\arabic{appen}.\arabic{equation}}

\newcounter{appen}
\newcommand{\appen}[1]{\par\refstepcounter{appen}
{\par\medskip\noindent\Large\bf Appendix \arabic{appen}. \medskip }{\bf \large{#1}}}

\renewcommand{\proposition}{\par\refstepcounter{proposition}
{\bf Proposition. }}
\makeatletter \@addtoreset{proposition}{appen}
\renewcommand\theproposition{A\theappen.\arabic{proposition}}

\renewcommand{\subsection}[1]{\refstepcounter{subsection}
{\bf A\arabic{appen}.\arabic{subsection}. }{\ \bf #1}}
\renewcommand\thesubsection{A\theappen.\arabic{subsection}}
\makeatletter \@addtoreset{subsection}{appen}

\renewcommand{\subsubsection}{\par\refstepcounter{subsubsection}
{\bf A\arabic{appen}.\arabic{subsection}.\arabic{subsubsection}. }}
\renewcommand\thesubsubsection{A\theappen.\arabic{subsection}.\arabic{subsubsection}}
\makeatletter \@addtoreset{subsubsection}{subsection}


\appen {$I_{1,5}$.}
\label{AppI}

Let $f=e^{zp}$, $g=e^{zq}$, and $h=e^{zr}$ in some vicinity of the point $x\in \R^2$,
$z=(x,\xi^1,\xi^2)$.

Because $I_{1,5}$ is local form, we can introduce the polynomial
\ben
I(p,q,r)\stackrel {def} = I_{1,5}(z|f,g,h)e^{-z(p+q+r)}.
\een

$I(p,q,r)$ is antisymmetric (in
usual, non"super" sense) 8-th - order polynomial of 3 variables, each
consists of 1 pair of even and 1 pair of odd components.

Obviously,
\begin{eqnarray*}
\{e^{zp},e^{zq}\} &=&[p,q]e^{z(p+q)},
\end{eqnarray*}
where
\begin{eqnarray*}
\lbrack p,q] & \stackrel {def} =  &\lbrack p,q]_x+\lbrack p,q]_\xi,\\
\lbrack p,q]_x & \stackrel {def} =  &p_{1}q_{2}-p_{2}q_{1},\\
\lbrack p,q]_\xi & \stackrel {def} =  &-p_{3}q_{3}-p_{4}q_{4}.
\end{eqnarray*}

Direct calculation gives the following form for $I(p,q,r)$:

\begin{equation*}
I=I_1
+I_{2}+I_{3}+I_{4}+I_{5}+I_{6}+I_7+I_8,
\end{equation*}
where
\begin{eqnarray*}
I_{1}(p,q,r)&=&([p,q+r]^{3}+[p+r,q]^{3}-[p,q]^{3})r_{4}r_{3}+
[p,q]^{3}(p+q)_{4}(p+q)_{3}+\mathrm{cycle}(p,q,r),
\\
I_{2}(p,q,r)&=&-6r_{4}r_{3}[p,q]^{2}[p,q]_{x}+\mathrm{cycle}(p,q,r),
\\
I_{3}(p,q,r)&=& 2r_{4}r_{3}
\{6[p,q](p_{2}q_{1}q_{2}+p_{1}p_{2}q_{2})r_{1}-
3[p,q](p_{2}q_{1}^{2}+p_{1}^{2}q_{2})r_{2}+
\\
&&+(p_{2}^{3}q_{1}-p_{1}q_{2}^{3}+3p_{2}q_{1}q_{2}^{2}
-3p_{1}p_{2}^{2}q_{2})r_{1}^{2}+3(p_{1}^{2}p_{2}q_{2}-
p_{2}q_{1}^{2}q_{2})r_{1}r_{2}+
\\
&&+(p_{2}q_{1}^{3}-p_{1}^{3}q_{2})r_{2}^{2}\}+\mathrm{cycle}(p,q,r),
\\
I_{4}(p,q,r)&=&
(p_{4}+q_{4})(p_{3}+q_{3})[(q_{1}^{3}q_{2}^{2}-p_{1}^{3}p_{2}^{2}+
3p_{1}q_{1}^{2}q_{2}^{2}-3p_{1}^{2}p_{2}^{2}q_{1}+ \\
&&+ p_{1}^{3}p_{2}q_{2}-p_{2}q_{1}^{3}q_{2}+3p_{1}^{2}q_{1}q_{2}^{2}
-3p_{1}p_{2}^{2}q_{1}^{2}+3p_{1}^{2}p_{2}q_{1}q_{2}-3p_{1}p_{2}q_{1}^{2}q_{2}+
\\
&&+ p_{2}^{2}q_{1}^{3}-p_{1}^{3}q_{2}^{2})r_{2}+(p_{1}+q_{1})^{2}(p_{2}^{3}
-q_{2}^{3})r_{1}]+ \mathrm{cycle}(p,q,r),
\\
I_{5}(p,q,r)&=&(p_{4}+q_{4})(p_{3}+q_{3})[(p_{1}+q_{1})^{2}(q_{1}-p_{1})r_{2}^{3}+
\\
&&+3(p_{1}+q_{1})^{2}(p_{2}-q_{2})r_{1}r_{2}^{2}+3(p_{1}
+q_{1})(q_{2}^{2}-p_{2}^{2})r_{1}^{2}r_{2}+ \\
&&\,+(p_{2}+q_{2})(p_{2}^{2}-q_{2}^{2})r_{1}^{3}]+\mathrm{cycle}(p,q,r),
\\
I_6(p,q,r) &=&-(p_4 +q_4 +r_4 )(p_3 +q_3 +r_3 )[p,q]^{3}+\mathrm{cycle}(p,q,r),
\\
I_7(p,q,r)&=&
(p_4 +q_4 +r_4 )(p_3 +q_3 +r_3 )(p_{1}+q_{1}+r_{1})^{2}\{[p,q](3(p_{2}+q_{2})r_{2}- \\
&&-(p_{2}+q_{2}+r_{2})^{2})\}+\mathrm{cycle}(p,q,r),
\\
I_8(p,q,r)
&=&
(p_4 +q_4 +r_4 )(p_3 +q_3 +r_3 )
\{(4p_{2}q_{2}r_{1}^{2}-2(p_{2}+
q_{2})r_{1}^{2}r_{2}+\\
&& +4(p_{1}+q_{1})r_{1}r_{2}^{2}+2r_{1}^{2}r_{2}^{2})[p,q]+
\\
&&+(2(p_{2}^{2}+q_{2}^{2})r_{1}^{2}+4p_{1}q_{1}r_{2}^{2}-(2p_{1}p_{2}+4p_{1}q_{2}+
\\
&&+4p_{2}q_{1}+2q_{1}q_{2})r_{1}r_{2})[p,q]_{\xi }\}+\mathrm{cycle}(p,q,r).
\end{eqnarray*}

\appen {The proof of Proposition \ref{prop2}.}
\label{App2}

\subsection{Low degree filtrations}

The following low degree filtrations $\oP_{p_1,p_2,...,p_s}$ of the
polynomials we define as:
\definition
\begin{eqnarray}
&&\oP_{p_1,p_2,...,p_s} = \{f(k_1,k_2,...,k_s)\in \K [k_1,...,k_s]:
\nonumber\\ && \ \ \ \ \ \ \exists g \in \K [\alpha_1,\alpha_2,...,\alpha_s,
k_1,...,k_s]\ \ f(\alpha_1 k_1,\alpha_2 k_2,...) = \alpha_1^{p_1}
\alpha_2^{p_2}...\alpha_s^{p_s} g\},
\end{eqnarray}
where $k_1$, $k_2$, ... ,
$k_s$ are some sets of supervariables, and $\alpha_1$, ... , $\alpha_s$ are
some even variables.

Evidently, $\oP_{p_1,p_2...} \subset \oP_{q_1,q_2...}$ if $p_i \geq q_i$.  It
is clear also, that if $f\in \oP_{p_1,p_2,...}$ and $g\in \oP_{q_1,q_2,...}$
then $fg\in \oP_{p_1 + q_1,p_2 + q_2,...}$.

\subsection{The proof of Proposition \ref{prop2}}

\proposition
{\it If $\alpha I_{1,5}=d_2^{\mathrm {ad}}S$,
$S\in {\cal C}_2(\oD, \mathbf E)$, then $\alpha=0$.}

{\bf Proof.}
Let us suppose that there exists such bilinear form $S$ that
\be\label{suppose}
\alpha I_{1,5} = d_2^{\mathrm {ad}} S.
\ee
Consider Eq. (\ref{suppose}) in the domain
${\cal U}\subset \R^2\times\oD\times\oD\times\oD$, where
$(x,f,g,h)\in{\cal U}$ if there exists such vicinity $V_x$ of $x$ in $\R^2$, that at least
one of the following relations is true:
\begin{eqnarray*}
&&
V_x\cap \supp f = \varnothing,\ \ \
V_x\cap \supp g = \varnothing,\ \ \
V_x\cap \supp h = \varnothing,\ \ \ \\
&&
\supp f\cap \supp g = \varnothing,\ \ \
\supp g\cap \supp h = \varnothing,\ \ \
\supp h\cap \supp f = \varnothing.\ \ \
\end{eqnarray*}
Here, as everywhere in present paper, $x$ is even coordinates of $z$:
$z=(x,\xi^1,\xi^2)$.

In this domain, $\hat {I}_{1,5}=0$, and so
$
d_2^{\mathrm {ad}} \hat S=0.
$
In such a way (see \cite{Cohom} for details)
$S$ has the form
$S=m_\zeta+c_3m_3+c_5m_5+d_1^{\mathrm {ad}}b^D+S_{\mathrm {loc}},$
where
$
S_{\mathrm {loc}}(z|f,g)=
\sum_{k,l=0}^{N}m^{(A)_{k}|(B)_{l}}(z)(\partial_{A})^{k}f(z)(\partial_{B})^{l}g(z).
$
Thus Eq. (\ref{suppose}) reduces to
\be\label{suppred}
d_2^{\mathrm {ad}}S_{\mathrm {loc}}=\alpha I_{1,5}.
\ee

Let $f=e^{zp}$, $g=e^{zq}$ and $h=e^{zr}$ in some vicinity of the point $x$.
Because $S_{\mathrm {loc}}$ is local form, we can introduce the polynomial
\ben
F(z|p,q)\stackrel {def} = S_{\mathrm {loc}}(z|e^{zp},e^{zq})e^{-z(p+q)}=-F(z|q,p).
\een

Then, the equation (\ref{suppred}) acquires the form
\be
 [p,q]\Phi (z|p,q,r)+\{F(z|q,r),zp\}+\mathrm{cycle}(p,q,r)
 \ =\alpha I(p,q,r),  \label{6.1.2}
\ee
where
\begin{equation*}
\Phi (z|p,q,r)=F(z|p+q,r)-F(z|p,r)-F(z|q,r)=\Phi (z|q,p,r).
\end{equation*}

The function $I(p,q,r)$ can be represented in the form
\begin{eqnarray*}
&&I(p,q,r)=\{[p,q]\varphi _{0}(p,q,r)+\mathrm{cycle}(p,q,r)\}+I_{(1)}, \\
&&I_{(1)}(p,q,r)=\{[p,q]r_{1}^{2}r_{2}^{2}r_{3}r_{4}+\mathrm{cycle}
(p,q,r)\}+\{p_{A}\varphi ^{A}(q,r)+\mathrm{cycle}(p,q,r)\}+ \\
&&+I_{(2)}(p,q,r),
\end{eqnarray*}
where
\begin{eqnarray*}
\varphi _{0}(p,q,r) &=&\varphi _{0}(p+q,r)-\varphi _{0}(p,r)-\varphi
_{0}(q,r), \\
\varphi _{0}(p,q)
&=&p_{3}p_{4}p_{1}^{2}p_{2}^{2}-q_{3}q_{4}q_{1}^{2}q_{2}^{2}, \\
\varphi ^{A}(p,q) &\in &\oP_{2,2},\;I_{(2)}(p,q,r)\in \oP_{2,2,2}.
\end{eqnarray*}
Further, represent $F(z|p,q)$ and $\Phi (z|p,q,r)$ in the form
\begin{eqnarray*}
&&F(z|p,q)=\varphi _{0}(p,q)+F_{0}(z|p,q)+F_{1}(z|p,q), \\
&&\Phi (z|p,q,r)=\varphi _{0}(p,q,r)+\Phi _{0}(z|p,q,r)+\Phi _{1}(z|p,q,r),
\\
&&F_{0}(z|p,q)=G(z|p)-G(z|q),\;G(z|p)\in \oP_1,\;F_{1}(z|p,q)\in \oP_{1,1}, \\
&&\Phi _{0}(z|p,q,r)=G(z|p+q)-G(z|p)-G(z|q)+G(z|r), \\
&&\Phi _{1}(z|p,q,r)=F_{1}(z|p+q,r)-F_{1}(z|p,r)-F_{1}(z|q,r).
\end{eqnarray*}
Then we have
\begin{eqnarray}
&&[p,q]\{\Phi _{0}(z|p,q,r)+\Phi _{1}(z|p,q,r)\}+\{F_{0}(z|p,q),zr)+  \notag
\\
&&\,+\{F_{1}(z|p,q),zr\}+\mathrm{cycle}(p,q,r)=\alpha I_{(1)}(p,q,r).
\label{A2.7}
\end{eqnarray}

Consider Eq. (\ref{A2.7}) for $r=0$:
\begin{eqnarray*}
&&[p,q]\{G(z|p+q)-G(z|p)-G(z|q)\}+ \\
&&\,+\{G(z|q),zp\}-\{G(z|p),zq\}=0.
\end{eqnarray*}
According to [4],\ it follows from this equation that $F_{0}(z|p,q)$ is
equal to coboundary up to term $\in \oP_{1,1}$ which we include in $F_{1}(z|p,q)
$. So, we obtain
\begin{equation}
\lbrack p,q]\Phi _{1}(z|p,q,r)\}+\{F_{1}(z|p,q),zr\}+\mathrm{cycle}
(p,q,r)= \alpha I_{(1)}(p,q,r).  \label{A2.8}
\end{equation}
$F_{1}(z|p,q)$ is a polynomial of some finite order $Q$, $I_{(1)}(p,q,r)$ is
a homogeneous polynomial of the eighth order. Let $F_{1(Q)}(z|p,q)$ be the
terms of the $Q$-th order. For $Q\geq 7$ we have
\begin{equation*}
\lbrack p,q]\Phi _{1(Q)}(z|p,q,r)+\mathrm{cycle}(p,q,r)=0.
\end{equation*}
According to [4],\ it follows from this equation that $F_{1(Q)}(z|p,q)$ is
equal to coboundary up to terms of orders $\leq Q-1$. So we find that $
F_{1}(z|p,q)=F_{1(6)}(z|p,q)$ up to coboundary and terms of orders $\leq 5$.
Thus we obtain from Eq. (\ref{A2.8})
\begin{equation}
\lbrack p,q]\Phi _{1(6)}(z|p,q,r)\}+\mathrm{cycle}(p,q,r)=
\alpha I_{(1)}(p,q,r).  \label{A2.10}
\end{equation}

Consider the terms of the first order in $r$ in Eq. (\ref{A2.10}):

\begin{eqnarray}
&&[p,q][F_{1(5)}^{A}(z|p+q)-F_{1(5)}^{A}(z|p)-F_{1(5)}^{A}(z|q)]=
-\alpha I_{(1)}^{A}(p,q),  \label{A2.11} \\
&&F_{1}^{A}(z|p)=\left. F_{1}(z|p,r)\frac{\overleftarrow{\partial }}{
\partial r_{A}}\right| _{r=0},\;I_{(1)}^{A}(z|p,q)=\left. I_{(1)}(p,q,r)
\frac{\overleftarrow{\partial }}{\partial r_{A}}\right| _{r=0}.  \notag
\end{eqnarray}

In Eq. (\ref{A2.11}), consider the terms of the type $pq^{6}$ and $qp^{6}$.
In this case, the l.h.s. of Eq. (\ref{A2.11}) equals to zero and we obtain
\begin{equation*}
0=\alpha (q_{A}p_{1}^{2}p_{2}^{2}p_{4}p_{3}-p_{A}q_{1}^{2}q_{2}^{2}q_{4}q_{3}).
\end{equation*}
So, $\alpha=0$.
\hskip 5mm
$\blacksquare$


\end{document}